\documentclass[letter,onecolumn]{jpsj3}
\usepackage{txfonts}
\title{Compressed Sensing in Scanning Tunneling Microscopy/Spectroscopy for Observation of Quasi-Particle Interference}

\author{Yoshinori~Nakanishi-Ohno$^{1}$\thanks{equally contributed}, Masahiro~Haze$^{2\,*}$, Yasuo~Yoshida$^2$, Koji~Hukushima$^{1,3}$, Yukio~Hasegawa$^2$, and Masato~Okada$^{4\, \dagger}$}
\inst{
$^1$Graduate School of Arts and Sciences, The University of Tokyo, Meguro, Tokyo 153-8902, Japan \\
$^2$The Institute for Solid State Physics, The University of Tokyo, Kashiwa, Chiba 277-8581, Japan \\
$^3$Center for Materials Research by Information Integration, National Institute for Materials Science, Tsukuba, Ibaraki 305-0047, Japan \\
$^4$Graduate School of Frontier Sciences, The University of Tokyo, Kashiwa, Chiba 277-8561, Japan}

\abst{
We applied a method of compressed sensing to the observation of quasi-particle interference (QPI) by scanning tunneling microscopy/spectroscopy to improve efficiency and save measurement time. 
To solve an ill-posed problem owing to the scarcity of data, the compressed sensing utilizes the sparseness of QPI patterns in momentum space. 
We examined the performance of a sparsity-inducing algorithm called least absolute shrinkage and selection operator (LASSO), 
and demonstrated that LASSO enables us to recover a double-circle QPI pattern of the Ag(111) surface from a dataset whose size is less than that necessary for the conventional Fourier transformation method. 
In addition, the smallest number of data required for the recovery is discussed on the basis of cross validation.
}

\begin{document}
\maketitle
The interference of electrons is one of the manifestations of their particle-wave duality in quantum mechanics.
When electrons are scattered by local disordered structures, such as defects, adsorbates, and step edges on surfaces, the reflected electronic wave interferes with the injected one to form a spatial modulation in the local density of states (LDOS), that is, a quasi-particle interference (QPI) pattern.
The modulated LDOS was first observed in real space by scanning tunneling microscopy and spectroscopy (STM/S) on noble metal surfaces. \cite{Hasegawa93,Crommie93}
Analyzing a QPI pattern provides us with much information on the electronic states in momentum ({\bf k}) space because the wavenumber of QPI patterns corresponds to the momentum difference ({\bf q}) between the reflected and injected electrons.
In addition, the energy dispersion relation of the states can be revealed from a stack of QPI patterns obtained at various bias voltages.
Recently, QPI analysis has been applied to investigate electronic/spin structures of complex materials such as cuprate superconductors \cite{Hoffman02,McElroy03,Hanaguri10} and topological insulators. \cite{Roushan09,Zhang09,Alpichshev10}
QPI observation with STM/S is thus expected to be a fundamental tool in various aspects of solid-state physics.

Two-dimensional (2D) STS is useful for imaging the energy dependence of QPI patterns.
In this method, the spectrum of tunneling conductance ($dI/dV$), which corresponds to the LDOS, is taken at every pixel while scanning, and then LDOS maps are made at various bias voltages.
The band structure of electronic states can be revealed by performing the Fourier transformation (FT) of the $dI/dV$ maps.
However, 2D STS takes quite a long time, which sometimes amounts to more than a week.
This means that high mechanical and thermal stabilities of the system are mandatory for the measurement, which makes the QPI observation difficult to perform.
The long time is attributed to the large number of $dI/dV$ spectra needed to obtain sufficient {\bf k}-space information; to cover the entire Brillouin zone, a tunneling spectrum has to be taken at every unit cell, and to improve the {\bf k}-space resolution, a wide real-space area has to be probed.
If one can obtain the same quality of {\bf k}-space information from a reduced number of spectra, the QPI analysis with STM/S will be much more convenient and widely used.

As a solution to the problem posed above, we apply compressed sensing (CS) to STM/S for the QPI observation.
CS is a novel statistical method for acquiring and reconstructing a signal efficiently, developed in the field of signal processing \cite{Donoho06a,Candes08}.
Elegant results of CS were reported in various fields of natural science such as magnetic resonance imaging in medical science,\cite{Lustig07,Lustig08} NMR in protein science,\cite{Kazimierczuk11,Holland11} and radio interferometry in astronomy.\cite{Honma14}
In the case of QPI observation, if the number of $dI/dV$ spectra is decreased, 
the FT of the $dI/dV$ map becomes of too low quality to access {\bf k}-space information on the electronic states.
If the resolution of the {\bf k} space is unreasonably enhanced, we will be confronted with an underdetermined problem, in which the number of variables measured is smaller than the number of variables to be determined.
CS addresses this problem by utilizing the sparseness of QPI patterns; their FTs are composed of few nonzeros and many zeros.
This sparseness is based on the fact that LDOS modulations with a small number of wavelengths are possible at a given energy in accordance with the energy dispersion relation.
Because of this sparseness, we can reduce the number of unknown variables significantly,
and therefore, we can obtain an FT of sufficient quality even with scarce data.

In this paper, we demonstrate that CS performs well by numerical simulations on a QPI pattern observed in a $dI/dV$ map of a Ag(111) surface.
The surface state of Ag(111) is described using a free-electron-like model, and the FT of the QPI pattern has a circular pattern whose radius corresponds to twice the wavenumber of the states.
We use an analysis method of CS called least absolute shrinkage and selection operator (LASSO) \cite{Tibshirani96} to recover the pattern from scarce data.
Then, we discuss by how much the number of the sampling data can be reduced from a statistical viewpoint.

Let us formulate the measurement process.
A $dI/dV$ map of $M$ pixels is composed of $M$ pairs of $\{\mathbf{r}_\mu,g_\mu\}$ ($\mu=1,2,\dots, M$), where $\mathbf{r}_\mu$ is the spatial coordinates of the $\mu$th measurement point and $g_\mu$ is the $dI/dV$ value corresponding to the LDOS at $\mathbf{r}_\mu$.
In this study, we focus on $dI/dV$ image data at a given bias voltage, but our discussion is valid for $dI/dV$ maps at any bias voltage.
Data points denoted by $\mathbf{r}_\mu$ usually form a square lattice on the sample surface but do not necessarily have to.
When the FT of the $dI/dV$ map is denoted by $f({\bf q})$, each of the $dI/dV$ values is represented by 
\begin{eqnarray}
	g_\mu=\frac{1}{2\pi}\int d\mathbf{q}f(\mathbf{q})e^{i\mathbf{q}\cdot\mathbf{r}_\mu}.\label{eq:IFT}
\end{eqnarray}
For convenience of analysis, the integral of $\mathbf{q}$ is approximated as the sum of integrand values at a large number of points $\mathbf{q}_j$ ($j=1,2,\dots, N$) forming a square lattice.
This discretization can be made as fine as needed at the expense of computational cost.
Then, $M$ discretized linear equations are obtained as
\begin{eqnarray}
	g_\mu=\sum_{j=1}^NG_{\mu j}f_j,\label{eq:DIFT}
\end{eqnarray}
where $f_j=f(\mathbf{q}_j)$ and $G_{\mu j}=\frac{1}{\sqrt{N}}e^{i\mathbf{q}_j\cdot\mathbf{r}_\mu}$.
The $M$-by-$N$ matrix $\mathbf{G}$ is a submatrix of the unitary matrix representing the inverse discrete FT.
The purpose of QPI observation is to estimate the $N$-dimensional vector $\mathbf{f}$ using the $M$-dimensional vector $\mathbf{g}$ and the $M$ equations of Eq. (\ref{eq:DIFT}).
If there are a sufficient amount of data, $M>N$, the conventional FT method yields an estimator $\hat{\mathbf{f}}_{\rm conv}$ of $\mathbf{f}$ given by
\begin{eqnarray}
	\hat{\mathbf{f}}_{\rm conv}=\mathbf{F}\mathbf{g},\label{eq:PI}
\end{eqnarray} 
where $\mathbf{F}$ is the Moore--Penrose pseudoinverse matrix of $\mathbf{G}$.

CS deals with the case where the number of equations $M$ is smaller than the number of variables $N$ to be determined ($M<N$).
Owing to the problem being underdetermined, there is no unique solution $\mathbf{f}$ that satisfies all of the equations in Eq. (\ref{eq:DIFT}), even without measurement noise.
Here, we assume that the FT of the $dI/dV$ image, denoted by $\mathbf{f}$, has sparseness.
Then, we only have to determine a small number of variables denoted by $K$.
It is expected that if the FT is sparse, it will be recovered from scarce measurement data because the problem is considered to be substantially overdetermined, namely, $M>K$.

In principle, the sparse solution is obtained by solving the following problem of $l_0$-norm minimization:
\begin{eqnarray}
	\hat{\mathbf{f}}_{l_0}=\mathrm{arg}\min_{\mathbf{f}}||\mathbf{f}||_0\ \textrm{subject to }\mathbf{g}=\mathbf{Gf},\label{eq:l0min}
\end{eqnarray}
where $||\cdot||_0$ represents the number of nonzero components and is called the $l_0$-norm.
However, the direct execution of the $l_0$-norm minimization is of little practical use because this optimization problem is classified as NP-hard (non-deterministic polynomial-time hard) in computational complexity theory.\cite{Natarajan95}
Putting it simply, the $l_0$-norm minimization involves checking whether each of the Fourier components $f_j$ is nonzero or zero, one by one, and its computational cost amounts to the order of $O(2^N)$.
In addition, the constraint of the right-hand side of Eq. (\ref{eq:l0min}) is too strong when it is used in the presence of measurement noise.
To cope with the difficulties, we therefore adopt the CS technique called LASSO.\cite{Tibshirani96}

In this paper, the LASSO estimator $\hat{\mathbf{f}}_{\rm LASSO}$ is defined by
\begin{eqnarray}
	\hat{\mathbf{f}}_{\rm LASSO}=\mathrm{arg}\min_{\mathbf{f}}\left\{\frac{1}{2}||\mathbf{g}-\mathbf{G}\mathbf{f}||_2^2+\lambda||\mathbf{f}||_1\right\},\label{eq:LASSO}
\end{eqnarray}
where $||\cdot||_2$ is the $l_2$-norm, also known as the Euclidean norm, and $||\cdot||_1$ is the $l_1$-norm, which represents the sum of the absolute values of the elements, namely, $||\mathbf{f}||_1=\sum_j|f_j|$.
The $l_2$-norm term is a data-fitting term and the $l_1$-norm term is a sparsity-inducing term.
The regularization coefficient $\lambda$, which controls the degree of sparseness, is often determined by the cross-validation (CV) technique.
The procedure of CV is as follows: 
i) divide a given data set ${\bf g}$ into two groups randomly, a training group ${\bf g}^{\rm train}$ and a testing group ${\bf g}^{\rm test}$; 
ii) obtain an estimate $\hat{\bf f}^{\rm train}$ by applying an analysis method to only the training data ${\bf g}^{\rm train}$;
iii) calculate the CV error (CVE) defined as the following mean-squared error,
\begin{eqnarray}
	{\rm CVE}=\frac{1}{2M^{\rm test}}\|{\bf g}^{\rm test}-{\bf G}^{\rm test}\hat{\bf f}^{\rm train}\|_2^2,\label{CVE}
\end{eqnarray}
where $M^{\rm test}$ is the dimension of ${\bf g}^{\rm test}$, and ${\bf G}^{\rm test}$ is a submatrix of {\bf G} whose rows correspond to ${\bf g}^{\rm test}$.
$k$-fold CV is often used to deal with the fluctuation of the CVE depending on the way the data are divided.
In $k$-fold CV, data are divided into $k$ equal-sized groups, and the procedure of CV is repeated $k$ times with each of the $k$ groups used exactly once as testing data.
The $k$ values of the estimated CVE are then averaged to yield a single estimate of the CVE.
As shown in Eq. (\ref{CVE}), the CVE measures how well the testing group ${\bf g}^{\rm test}$ is described by the estimate $\hat{\bf f}^{\rm train}$; the smaller the CVE, the better the method performs.
Therefore, CV is useful for evaluating the performance of analysis methods and for setting the regularization parameter $\lambda$ in LASSO to an appropriate value.

Let us briefly explain the derivation of LASSO.
Technically, Eq. (\ref{eq:LASSO}) is the Lagrangian form of LASSO, also known as basis pursuit denoising (BPDN),\cite{Chen98} but the essence of LASSO still remains in the $l_1$-norm term.
In the first place, basis pursuit (BP) refers to a problem defined by replacing the $l_0$-norm in Eq. (\ref{eq:l0min}) with the $l_1$-norm.
It is well known that BP can be recast as a linear program.\cite{Bloomfield83}
Recently, it has been mathematically demonstrated that, under reasonable conditions, the solution of Eq. (\ref{eq:l0min}) can be obtained by solving BP.\cite{Donoho05a,Donoho05b,Candes05,Candes06,Candes07}
To handle the noisy case, it was argued that the constraint of BP is made weaker by replacing it with an inequality constraint $||\mathbf{g}-\mathbf{G}\mathbf{f}||_2^2<\epsilon$, where $\epsilon$ is a small positive number.\cite{Donoho06b}
Actually, the original LASSO was introduced as the minimization of the squared error $||\mathbf{g}-\mathbf{G}\mathbf{f}||_2^2$ with an inequality constraint $||\mathbf{f}||_1<\epsilon$.\cite{Tibshirani96}
However, inequality constraints are inconvenient for implementation, and equivalent Lagrangian forms such as Eq. (\ref{eq:LASSO}) are preferred.
There are some efficient algorithms for carrying out the minimization in Eq. (\ref{eq:LASSO}): least-angle regression,\cite{Efron04} approximate message passing,\cite{Donoho09,Donoho10} and dual augmented Lagrangian \cite{Tomioka11}.

\begin{figure}[!t]
\centering
\includegraphics[width=\linewidth]{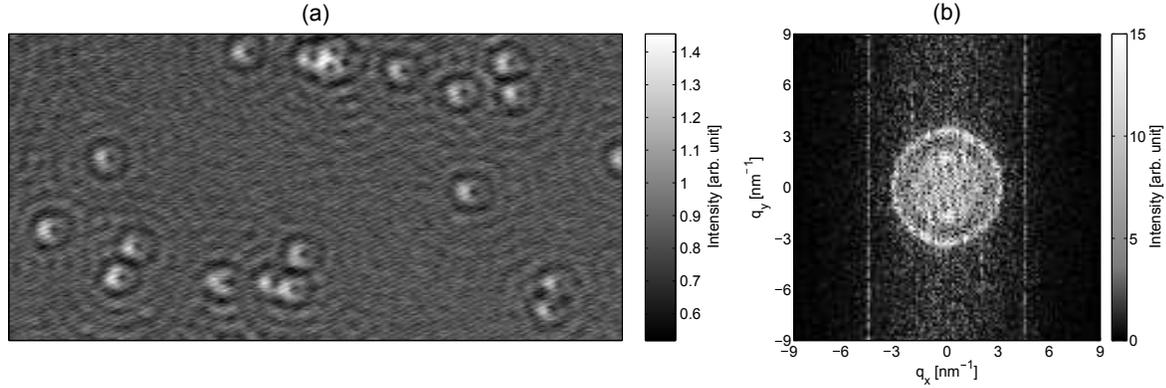}
\caption{
(a) $dI/dV$ map of Ag(111) surface.
(b) FT of (a) obtained by conventional method.
}
\label{fig1}
\end{figure}

We investigate the performance of CS by numerical simulations with actual data of $dI/dV$ mapping.
We use the $I/V$ map of the Ag(111) surface shown in Fig. \ref{fig1}(a).
The experiments were performed using an ultrahigh-vacuum STM setup (USM-1300, Unisoku, and SPM-1000, RHK) in which the tip and sample can be cooled to 2.6 K at the Institute for Solid State Physics, The University of Tokyo.
A single-crystalline Ag(111) substrate was cleaned by repeated Ar sputtering and annealing at 800 $^\circ$C.
The $dI/dV$ image of the Ag(111) surface was obtained at 4.2 K.
An electrochemically etched W tip, which was annealed at 900 $^\circ$C {\it in situ} to remove the oxide layer from the tip apex, was used for the imaging.
The image data were obtained at a sample bias voltage of 200 mV, indicating that the image corresponds to the LDOS mapping at 200 meV above the Fermi energy.
The size of the observed region is 70 $\times$ 35 nm$^2$.
The number of pixels is 360 $\times$ 180, namely, $M=64800$. 
There is a QPI pattern on the surface.
The FT of this $dI/dV$ map obtained by the conventional method is shown in Fig. \ref{fig1}(b).
The $\mathbf{q}$-space region of interest is discretized into 128 $\times$ 128 pixels, namely, $N=16384$.
In this case, the amount of data is sufficient compared with the number of unknown variables ($M>N$).
Then, we can see two rings centered at the origin in the $\mathbf{q}$ space as reported by Sessi {\it et al.} \cite{Sessi15}
The outer ring corresponds to the conventional surface electronic states and the inner one is due to the acoustic surface plasmon.
The region occupied by the two rings is small in the $\mathbf{q}$ space, and thus the sparseness assumption holds in this case.

\begin{figure}[!t]
\centering
\includegraphics[width=\linewidth]{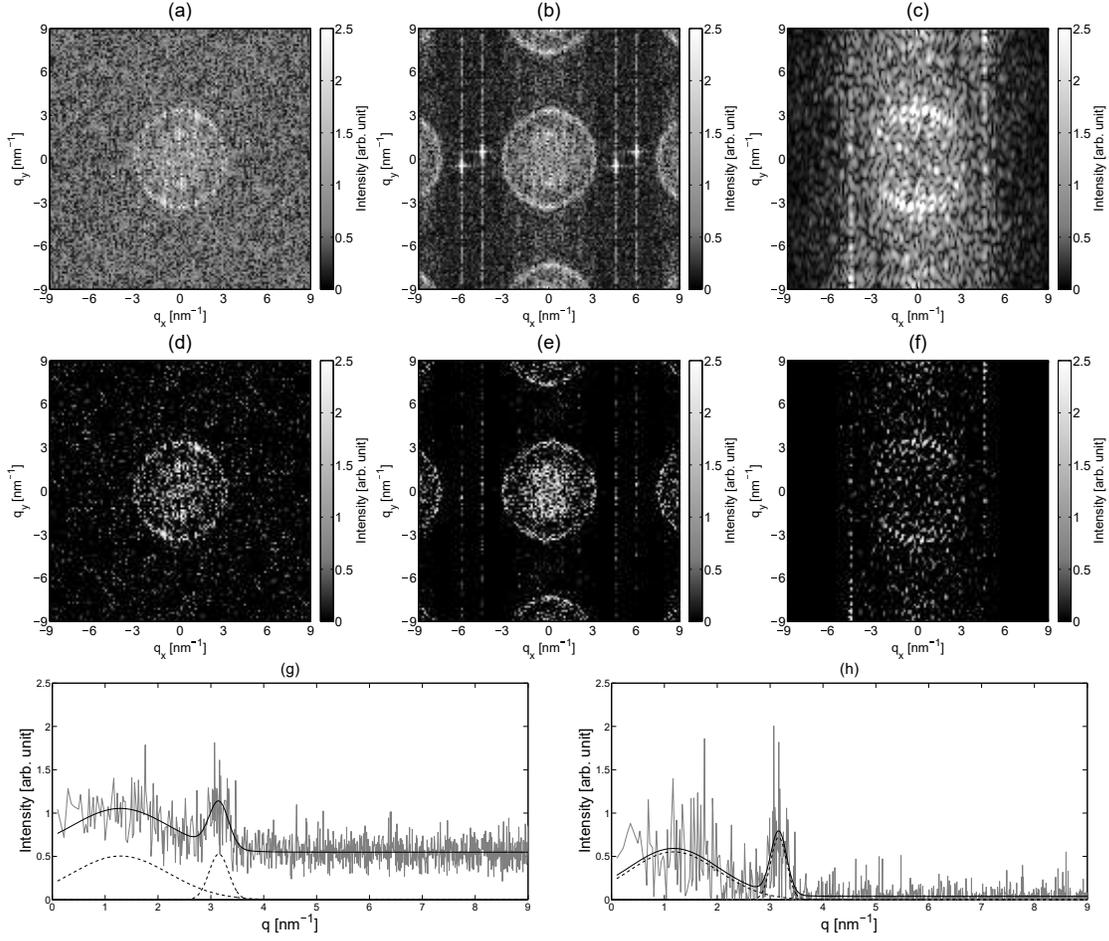}
\caption{
(a)--(f) FTs estimated from different parts of Fig. \ref{fig1}(a).
Subsets of data are composed of 7200 pixels.
Results of the conventional method and LASSO are shown in the top and middle rows, respectively.
(a) and (d) are obtained by using randomly chosen pixels.
(b) and (e) are obtained by using every third pixel in both the horizontal and vertical directions.
(c) and (f) are obtained by using only the central region in the ${\bf r}$ space.
(g) and (h) are radially averaged line sections corresponding to (a) and (d), respectively.
Each of the fitting curves is composed of two Gaussian functions and a background constant.
(g) $\mu_1=1.3\textrm{ nm}^{-1}$, $\mu_2=3.2\textrm{ nm}^{-1}$, $\sigma_1=0.91\textrm{ nm}^{-1}$, $\sigma_2=0.18\textrm{ nm}^{-1}$, $a_1/c=0.92$, $a_2/c=0.96$.
(h) $\mu_1=1.2\textrm{ nm}^{-1}$, $\mu_2=3.2\textrm{ nm}^{-1}$, $\sigma_1=0.84\textrm{ nm}^{-1}$, $\sigma_2=0.15\textrm{ nm}^{-1}$, $a_1/c=14$, $a_2/c=18$.
}
\label{fig2}
\end{figure}
Let us examine whether the double circle can be recovered from a reduced amount of data.
Figure \ref{fig2} shows FTs of $N=16384$, which are estimated from partial data of $M=7200$.
It is stressed that the number of unknown variables is larger than that of the measured variables.
The top and middle rows show the results of the conventional method and those of LASSO, respectively.
LASSO is carried out on the basis of approximate message passing. \cite{Donoho09,Donoho10}
The regularization parameter $\lambda$ of LASSO is set so as to minimize the CVE calculated by 10-fold CV.
The analysis with LASSO outperforms that by the conventional method in reducing background noise, and the expected pattern is more clearly seen.
In the ill-posed situation, LASSO provides a sparse solution; most of the noise components are automatically estimated to be zero and the signal components remain nonzero.

Figures \ref{fig2}(a) and \ref{fig2}(d) show the case of randomly chosen data points.
In this case, LASSO succeeds in recovering the double-circle pattern, whereas the conventional method fails.
Figures \ref{fig2}(b) and \ref{fig2}(e) show the case of every third data point in both the horizontal and vertical directions.
In both of these figures, we find phantom patterns around the true pattern.
These patterns are attributed to the aliasing effect that makes some different wavenumber components indistinguishable owing to the periodicity of sampling.
Figures \ref{fig2}(c) and \ref{fig2}(f) show the case of data points in a small central region of Fig. \ref{fig1}(a).
As shown in the figures, the QPI pattern is considerably deteriorated.
The inner ring disappears because long-wavelength components can hardly be detected in the small region.
In addition, the intensity of the outer ring is very weak because the region is too distant from defects to form the QPI pattern.
On the basis of the above results, we conclude that CS performs well with random sampling in a broad region.

To quantitatively evaluate the effects of LASSO, we examine line sections of the FTs.
Figures \ref{fig2}(g) and \ref{fig2}(h) show the radially averaged line sections of Figs. \ref{fig2}(a) and \ref{fig2}(d), respectively.
Each of the fitting curves is obtained by the least-squares method, where, in the same way as Sessi {\it et al.} \cite{Sessi15}, the following model function is employed: 
\begin{eqnarray}
	f(q)=\sum_{k=1,2}a_ke^{-\frac{1}{2\sigma_k^2}(q-\mu_k)^2}+c,
\end{eqnarray}
where $\{\mu_1,\mu_2,\sigma_1,\sigma_2,a_1,a_2,c\}$ is a set of model parameters.
Assume without loss of generality that $\mu_1<\mu_2$.
Note that the peak of the outer ring has a delta-function-like shape when the phase noise that arises from the random distribution of scatterers is removed.
The results show that LASSO provides a much higher signal-to-noise ratio ($a_k/c$) and determines the peak locations with less uncertainty ($\sigma_k$).
Overall, the synergy between measurement and analysis, namely, random sampling and LASSO, is indispensable for compressed sensing.

\begin{figure*}[!t]
\centering
\includegraphics[width=\linewidth]{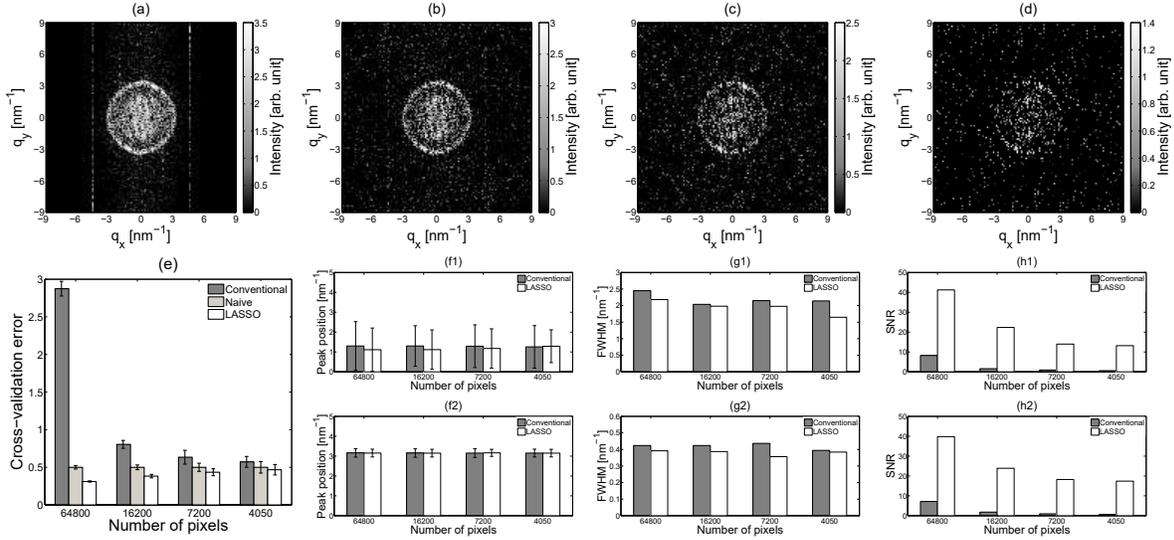}
\caption{
(a)--(d) FTs estimated from randomly reduced data with LASSO.
The number of pixels is $M=64800$, 16200, 7200, and 4050 from the left.
(e) Cross-validation error of conventional method, naive method, and LASSO.
The length of the error bars shows the standard deviation among 10 trials of 10-fold CV.
(f)--(h) Parameters of curve fitting to radially averaged line sections.
(f) Wavenumber of peak positions: (f1) $\mu_1$ and (f2) $\mu_2$.
The length of the error bars is the FWHM of the Gaussian functions shown in (g): (g1) $2\sqrt{2\ln2}\sigma_1$ and (g2) $2\sqrt{2\ln2}\sigma_2$.
(h) Signal-to-noise ratio (SNR): (h1) $a_1/c$ and (h2) $a_2/c$.
} 
\label{fig3}
\end{figure*}

Next, we discuss the performance of LASSO when it is applied to different amounts of data.
The top row of Fig. \ref{fig3} shows the FTs estimated from randomly reduced amounts of data.
In the cases of (a)--(c), the ring pattern is clearly seen as is expected, but in the last case of (d), we see that the ring pattern breaks at many places.
The data set of $M=4050$ is considered to be insufficient in quantity. 
This situation indicates that LASSO fails in the case of too scarce data.
However, it is difficult to judge whether the result in Fig. \ref{fig3}(d) is reliable when we do not know the true pattern in practice.
We argue that the CVE is a good criterion for evaluating the sufficiency of data.
Figure \ref{fig3}(e) shows the CVE of the conventional method, LASSO, and a naive method by which one blindly accepts the sparseness assumption and estimates that $\hat{\mathbf{f}}=\mathbf{0}$ without any concern about data fitting.
In the case of $M=64800$, LASSO gives a much smaller CVE than the conventional method.
The large CVE of the conventional method is attributed to overfitting to noise components in the data.
Although the naive method has a lower CVE than the conventional method, the naive method is still inferior to LASSO because it corresponds to applying LASSO with an infinitely large $\lambda$.
Here, let us focus on the fact that the difference in the CVE between the naive method and LASSO becomes smaller as the amount of data decreases.
To investigate the significance of the performance difference, we use an orthodox method of hypothesis testing called the $t$-test. 
When the $t$-test is used in natural science, the significance level is often set to $\alpha=0.01$.
According to the $t$-test at $\alpha=0.01$, a significant difference in the CVE remains when $M\geq 7200$ but not when $M=4050$.
This means, conversely, that the dataset of $M=4050$ is insufficient because even LASSO provides almost the same result as that of the naive method.
Consequently, it turns out that the minimum amount of data required for CS is between $M=4050$ and 7200.

To make progress on the methodology of CS, we discuss the use of other prior knowledge in addition to the sparseness assumption.
Let us remind ourselves that the surface state of Ag(111) is described by a free-electron-like model and that the QPI pattern is isotropic.
Then, it is important to argue the results on the basis of radially averaged line sections.
As seen in Fig. \ref{fig3}(f), the peak wavenumbers in the radial direction can be estimated correctly both by the conventional method and by LASSO even when $M=4050$.
Then, the combined use of the sparseness and isotropy assumptions will enable more efficient measurement.
Figures \ref{fig3}(g) and \ref{fig3}(h) show the full width at half maximum (FWHM) of the Gaussian functions and signal-to-noise ratio, respectively.
LASSO provides a smaller FWHM and a higher signal-to-noise ratio and works better than the conventional method.
Under the combined assumption, LASSO should enable us to save more measurement time.
Moreover, there will probably be room to develop a novel method of measurement and analysis if one takes advantage of available knowledge on the system.

In conclusion, we applied CS to QPI observation by STM/S to save experimental time.
The key assumption is the sparseness of QPI patterns in the $\mathbf{q}$ space.
Our numerical simulations demonstrated that LASSO enables us to recover the double-circle pattern of Ag(111) from a randomly reduced dataset, with which the conventional FT method fails.
We also pointed out that CV is useful for identifying the minimum amount of data required.
Future work will confirm the performance of compressed sensing by putting it into practice.

\section*{Acknowledgments}
This work was supported by JSPS KAKENHI Grant Numbers 16H01534 (YY), 25120010 (KH), and 25120009 (MO).

\end{document}